\documentclass[preprint,5p]{elsarticle}

\usepackage{graphicx}
\usepackage{url}
\usepackage{booktabs}
\usepackage{multirow}
\usepackage{paralist}

\usepackage[usenames,dvipsnames]{color}
\usepackage{xspace}

\newcommand{\eg}{e.g.,\xspace}
\newcommand{\ie}{i.e.,\xspace}
\newcommand{\ergo}{\emph{ErGo}\xspace}
\newcommand{\SoftwareDev}{\texttt{SoftwareDev}\xspace}

\journal{Computers \& Industrial Engineering}

\begin{document}

\begin{frontmatter} 
	\title{Modelling Competences for Partner Selection in Service-Oriented Virtual Organization Breeding Environments\tnoteref{extendedNote}}
	\tnotetext[extendedNote]{This paper is an extended version of the conference paper that appeared as~\cite{paszkiewicz:cscwd:2011}.
The key additions of this journal version are as follows: Section 2 has been largely extended, a running example illustrates the presented solutions, and Sections~6 and 7 have been added.}
	

	\author[UEP]{Zbigniew Paszkiewicz}
\ead{zpasz@kti.ue.poznan.pl}

\author[UEP]{Willy Picard\corref{cor}}
\ead{picard@kti.ue.poznan.pl}

\address[UEP]{Pozna\'{n} University of Economics, 
Department of Information Technology, 
al.~Niepodleg{\l}o\'{s}ci 10, 61-875 Pozna\'{n}, 
Poland}

\cortext[cor]{Corresponding author}

	\begin{abstract}
In the context of globalization and dynamic markets, collaboration among organizations is a condition \textit{sine qua non } for organizations, especially small and medium enterprises, to remain competitive. Virtual organizations have been proposed as an organizational structure adapted to collaboration among organizations. The concept of Virtual Organization Breeding Environment (VOBE) has been proposed as a means to support the creation and operation of virtual organizations. With the rise of the service-oriented architecture (SOA), the concept of service-oriented VOBE (SOVOBE) has been proposed as a VOBE systematically organized around the concept of services. In the context of SOVOBEs, novel competence models supporting both service orientation and collaboration among organizations have to be developed to support efficiently partner selection, a key aspect of VO creation. In this paper, such a competence model is presented. Our competence model consists of a competence description model, a competence verification method, and a competence search method.
The competence description model is an information model to describe organizations, their competences, and services they provides.
The competence verification method enables the verification of the reliance and relevance of competence descriptions.
The competence search method allows a VO planner to select appropriate partners based on VO specifications, encompassing competence requirements. Finally, implementation concerns based on the development of the prototype \ergo system are presented.

\end{abstract}

\end{frontmatter} 

\section{Introduction}

In the context of globalization and dynamic markets, collaboration among enterprises allows them to face the socio-economical challenges related with high dynamism and ever changing clients' needs. To be competitive, organizations base their operation on strategies of specialization, differentiation and collaboration within networks of organizations, referred to as \emph{Virtual Organization (VO)}~\cite{book:lcm:2008}. The main challenge of VOs is an efficient collaboration of autonomous partners to achieve a  predefined goal, and, if needed, to quickly adapt to changing environment. Adaptation helps to reduce business risk or to take advantage of new business opportunities~\cite{article:paszkiewicz:imda:2011}.

The creation of a VO is one of the most important phases in VO lifecycle. A key issue is the selection of appropriate partners and their services that could achieve the VO goal. The selected set of partners and services has a direct impact on the VO efficiency, effectiveness, ability to be adaptive, and dynamics. A selection of inappropriate partners, based on limited, unreliable information, leads to inefficient collaboration among organizations and, in consequence, to VO failure.

The concept of \emph{Virtual Organization Breeding Environment(VOBE)} has been proposed as a means to ease and foster the creation and operations of VOs. A VOBE gathers organizations, referred to as VOBE members, and provide them with support for future and current collaboration within VOs. A VOBE facilitates the whole VO lifecycle by limiting the open universe of partners and services to those registered, it may impose a standardized approach to description of competences, services, used infrastructures, etc. on its members. A VOBE may also provide its members with services, \eg negotiation tools and partner selection services.

Service orientation is emerging at multiple organizational levels in business, in response to the growing need for greater business integration, flexibility, and agility. Ser\-vice-oriented architecture (SOA) has been suggested as a valuable approach for the architecture and implementations of VOBEs and integration of collaborating organizations~\cite{prove:rabelo:2008}. A VOBE implemented in the SOA paradigm is referred in this paper as a \emph{Service-Oriented Virtual Organization Breeding Environment (SOVOBE)}.

To promote itself and to be taken into account during VO partner selection processes, each SOVOBE member should provide detailed and up-to-date information about the activities it can perform and the services it can offer~\cite{article:ermilova:2010}. This description is often referred to as \emph{competence description}. A key element for agile collaborative enterprises is a sound computer support for competence management, providing tools for partner and service selection based on competence description.

In this paper, we argue that existing competence models are not adapted to organizations willing to collaborate within VOs in SOVOBEs. The two characteristic elements of VOs in SOVOBEs---service orientation and cross-organization collaboration---are not taken into account in most competence models, with a notable exception: the \mbox{4-C} model~\cite{article:ermilova:2010}. The 4-C model has been developed to model competences in VOs, but it lacks support for service orientation. Additionally, competence models usually concentrate on a data model for the description of organizations competences, neglecting the verification of these competence descriptions and their actual use during the VO creation process.

In this paper, we propose a novel competence model to support partner selection in the VO creation process in SOVOBEs. The main contributions of this paper are \begin{inparaenum}[\itshape a\upshape)] \item a competence description model, which may capture information about an organization profile, its competences, and the services it provides, \item a competence verification method to evaluate the reliability of the information in the competence description model, and \item a competence search method that has been integrated into a partner search method adapted to the characteristics of SOVOBEs, \ie service orientation and cross-organization collaboration.
\end{inparaenum}

%


The remainder of this paper is organized as follows. In Section~\ref{sec:researchBackground}, the research background is presented. First, the concepts of VO and VOBE are described in more detail. Next, the service orientation applied to organizations is discussed, leading to the definition of SOVOBE. Then, the concept of competence modelling in VOs ends this section. In Section~\ref{sec:introducingTheProposedCompetenceModel}, the rationale and a general overview of the proposed competence model and its components---the competence description model, the competence verification method, and the competence search method ---are presented. In Section~\ref{sec:competenceDescriptionModel}, the competence description model is detailed. The three parts of the competence description model are then described: the competence profile, the service business profile, and the organization profile. In Section~\ref{sec:competenceVerificationMethod}, the competence verification method is presented, with a special emphasis on the role of a social network in this method. In Section~\ref{sec:competenceSearchMethod}, the competence search method is presented in the context of the partner selection process, in particular as an element of the Multi-Aspect Partner and Service selection method. Section~\ref{sec:implementationConcerns} outlines the intended application of the model in the construction sector and presents implementation concerns related with the \ergo system, a prototype implementation of the proposed approach. Finally, Section~\ref{sec:conclusions} concludes the paper.

\section{Research background}
\label{sec:researchBackground}

In this section, concepts concerning the collaboration of organizations in VOs are introduced. A special emphasis is put on (1) the concept of SOVOBE, and (2) the description of competences of SOVOBE members.

\subsection{Service-Oriented Virtual Organization Breeding Environment}

In this section, concepts and definitions referring to VOs and their operation within a VOBE are presented. Next, Service-Oriented Architecture (SOA) at the coarser level of organizations is discussed.

\subsubsection{Virtual organizations}

Organization environment is defined as ``all the forces, processes and other entities---such as companies, public administration agencies, non-government organizations---outside an organization that interact with the organization and can potentially affect the organization's performance''~\cite{book:stoner:1995:management}. In a global economy, organization environment strongly influences organization's operation and its market success. Current economic trends---globalization, development and proliferation of information technology, development of knowledge-based economy and rising competition---result in increased complexity, uncertainty, dynamism, turbulence and diversity of organization environment. Such an environment is particularly challenging for small organizations. Although small organizations are flexible, innovative, and able to adapt to a changing environment in a relatively easy and rapid way, they have limited capabilities to influence the market, to control their environment, and, finally, to compete with large global organizations that have much more resources. To remain competitive, small organizations may base their operation on strategies of specialization, differentiation and collaboration~\cite{book:porter:1985}.

The concept of a \emph{Virtual Organization (VO)}~\cite{book:lcm:2008} has been proposed as an approach to support collaboration among multiple autonomous partners---organizations, humans or information systems---with the help of information technologies, telecommunication and computer networks. In this paper, `virtual organization' is to be understood as defined by Paszkiewicz and Cellary~\cite{article:paszkiewicz:imda:2011}:
\begin{quote}
a Virtual Organization (VO) is a set of at least two autonomous partners, where at least one of them is an organization, collaborating within a particular structure of social and legal relations in order to carry out a particular venture due to the demand from virtual organization clients and having a plan to carry out this venture.
\end{quote}
The main challenge of VOs are, first, an efficient collaboration of autonomous partners to achieve a predefined goal, and, second, quickly adaptation abilities to changing environments. Adaptation may help organization to reduce business risk and to take advantage of new business opportunities. An overview of the theoretical foundations for VOs may be found in~\cite{book:lcm:2008} and~\cite{article:lcm:CNO:2009}.

Efficient collaboration requires the use of appropriate \emph{management strategies, techniques and structures}, \eg outsourcing and out-tasking strategies, techniques for efficient control of activity execution, standardization of non-critical areas of operation, and interorganizational integration on various levels of organizational structure. Efficient collaboration also requires new approaches to traditional areas of management, such as evaluation of organization performance, as the evaluation of a set of collaborating organizations largely differs from the evaluation of a single organization. 

Besides the development of management strategies, techniques, and structures appropriate to VOs, \emph{information technology solutions} have been proposed to support VOs in the whole spectrum of their strategic, tactical, and operational activities, \eg efficient planning of operation, communication and coordination of actions, integration of partners, control of activity execution, and measurement of business effectiveness.

\subsubsection{Virtual Organization Breeding Environment}

The concept of Virtual Organization Breeding Environment (VOBE, sometimes referred in the literature as VBE) has been proposed to facilitate, among others, the VO creation process. In this paper, `VOBE' is to be understood as defined by Camarinha-Matos, Afsarmanesh, and Ollus~\cite{book:lcm:2008}: 
\begin{quote}
a VOBE is an association of organizations with the main goal of increasing preparedness of its members towards collaboration in potential VOs.
\end{quote}
Various types of VOBEs may be distinguished, such as start-up incubators, technology clusters, or industry areas~\cite{article:lcm:CNO:2009}.

A VOBE provides a set of tools and sources of information that may be used in the VO creation process, \eg competence repositories, negotiation tools, and history of collaboration. A VOBE allows potential collaborators to prepare their future collaboration with other VOBE members before a business opportunity occurs. A VOBE supports its members during the whole VO lifecycle~\cite{book:picard:2010:itsoa}:
\begin{itemize}
	\item	in the VO creation phase: a VOBE may provide its members with access to information not publicly available, such as information about the past performance of VOBE members. It may also provide a standardized description of partner profiles, competences and services. It may support the partner selection process. It may provide methods and tools for analysis and evaluation of present and future collaboration performance, as well as necessary information for trust building among selected members;
	\item	in the VO operation phase: a VOBE may support communication and exchange of documents. It may facilitate integration of heterogeneous information systems and manage common infrastructure. It may provide guidelines for standardized data formats, data storage facilities, information about changing environment (context) of collaboration, information about new collaboration opportunities. It may permit to reuse information elaborated by other VOs (\eg business process models, best practices);
	\item	in the VO evolution phase: a VOBE may support adaptation by providing tools for the redefinition of business goals, for the search of new partners, and by supporting negotiations;
	\item	in the VO dissolution phase: a VOBE may support knowledge inheritance, \ie it may capture experience gained during the operation of VOs for future reuse.
\end{itemize}

\subsubsection{Service-oriented organizations}

Service orientation is emerging at multiple organizational levels in business, 
in response to the growing need for greater business integration, flexibility, and agility:

\begin{quote}
service-oriented architecture (SOA) is a par\-a\-digm for organizing and utilizing distributed capabilities that may be under the control of different ownership domains. It provides a uniform means to offer, discover, interact with and use capabilities to produce desired effects consistent with measurable preconditions and expectations.~\cite{standards:OASIS:SOA:referenceModel}
\end{quote}

Although service-oriented architecture is mainly implemented with Web services, it should not be restricted to a given technology or a given technical infrastructure~\cite{book:picard:2010:itsoa, article:brittenham:2007}. Instead, it ``reflects a way of thinking about processes that reinforces the value of commoditization, reuse, semantics and information, and creates business value''~\cite{book:bieberstein:2005}. According to Spohrer, Maglio, Bailey, and Gruhl~\cite{article:spohrer:2007}, a service is ``the application of competence and knowledge to create value between providers and receivers. The value that accrues is derived from the interactions of entities that are known as service systems~\cite{article:vargo:2004}''. 

In this paper, `service' is to be understood as defined in~\cite{standards:OASIS:SOA:referenceArchitecture}: 
\begin{quote}
a service is a mechanism to enable access to one or more capabilities, where the access is provided using a prescribed interface and is exercised consistent with constraints and policies as specified by the service description.
\end{quote}


The world economy is currently in an advanced stage of transformation from a goods-based economy to an economy in which value creation, employment, and economic wealth depend on the service sector ~\cite{article:spohrer:2008}. Service-oriented thinking is one of the fastest growing paradigms in IT, with relevance to accounting, finance, supply chain management and operations, strategy and marketing. This fact is confirmed by the statistic data provided in~\cite{article:demirkan:2008}:
\begin{itemize}
 \item	in year 2004, services already accounted for 75\% of the US gross domestic product (GDP) and 80\% of the private sector employment~\cite{techreport:pal:2005};
 \item	services play a similarly important role in all of the Organization for Economic Cooperation and Development (OECD) countries~\cite{article:demirkan:2008};
 \item	industries that deliver consulting, experience, information, or other intellectual content in 2004 accounted for more than 70\% of total value added in these countries~\cite{presentation:spohrer:2005};
 \item	market-based services, excluding those provided by the public sector (\eg education, health care, and government) accounted for 50\% of the total value added in these countries, and have become the main driver of productivity and economic growth, especially as the use of IT services has grown \cite{presentation:spohrer:2005};
 \item	worldwide end-user spending on IT services is supposed to grow at a 6.4\% compound annual growth rate to reach US 855.6 billion, with positive growth in nearly all market segments~\cite{forecast:babaie:2006};
 \item companies that implement a service-oriented architecture are able to reduce costs for the integration of projects and maintenance by at least 30\%~\cite{www:wall:2007};
 \item at least one-third of business application software spending will be on software-as-a-service, instead of on product licenses by 2012~\cite{forecast:plummer:2008}.
\end{itemize}

\subsubsection{Service-Oriented VOBE}

The goal of service science is ``to catalog and understand service systems, and to apply that understanding to advancing our ability to design, improve, and scale service systems for practical business and societal purposes''~\cite{article:demirkan:2008}. In this context, SOA has been suggested as a valuable approach for the architecture and implementations of VOBEs and integration of collaborating organizations~\cite{prove:rabelo:2008}. A VOBE implemented in the SOA paradigm is referred in this paper as a \emph{Service-Oriented Virtual Organization Breeding Environment (SOVOBE)}. A SOVOBE is systematically organized around the concept of services, which is not limited to Web services, but which encompasses also services performed by humans and organizations. In this paper, only SOVOBEs are taken into account. The concept of SOVOBE has been introduced in~\cite{book:picard:2010:itsoa}. 

Concepts underlying SOA may be applied at the coarse level of organizations within the context of SOVOBEs:
\begin{itemize}
	\item	\emph{service reuse} - a given organization may provide and consume the same service within many VOs;
	\item	\emph{service abstraction} - the details of the implementation of services offered by a given organization within a VO are usually hidden for other organizations, because the implementation of the core business services is associated with the know-how capital that gives the organization a competitive advantage over other organizations;
	\item	\emph{service discoverability} - services provided by organizations in a SOVOBE are described so that both services and associated organizations may be identified as potential VO partners for a given business opportunity;
	\item	\emph{service composition} - a complex service provided by a VO is a result of the  composition of services provided by VO partners and eventually by the SOVOBE.
\end{itemize}

Depending on the type of service providers and consumers, the following classification of services provided by SOVOBEs has been proposed~\cite{book:picard:2010:itsoa}:
\begin{itemize}
 \item	\emph{business services} - services provided by SOVOBE members for chosen VO partners;
 \item	\emph{internal services} - services provided by the SOVOBE and consumed by its members. This set of services includes services for partner selection, process modelling tools adapted to VOs, social network modelling, performance estimation, and competence modelling;
 \item	\emph{external services} - services provided by organizations operating outside the SOVOBE, but offered through the SOVOBE to its members. External services facilitate interactions between external organizations (\eg public administration units) and the SOVOBE, its members and the VOs it has bred;
 \item	\emph{fa\c{c}ade services} - services provided by the SOVOBE to organizations outside the SOVOBE. Fa\c{c}ade services provide external organizations with access to information about the SOVOBE, such as its  members' profiles. Fa\c{c}ade services also allow external organizations to submit information to the SOVOBE, such as announcements of business needs.
\end{itemize}

\subsection{Competence Modelling}

To promote itself and to be taken into account during VO partner selection processes, each SOVOBE member should provide detailed and up-to-date information about the activities it can perform and the services it can offer. This information should be ``an accurate description of member capabilities, its free resources capabilities, the production costs for each of its products, as well as conspicuous proof of the validity of the provided information`` ~\cite{article:ermilova:2010}, usually referred to as \emph{competences}.

Various definitions of the concept of \emph{competence} have been proposed in the literature: in \cite{prove:ermilova:2006}, competence is defined as ``the organization's capability to perform (business) processes, tasks, having the necessary resources (human, technological, physical) available, and applying certain standards (practices), with the aim to offer certain products and/or services''. In~\cite{article:sanchez:1997}, competence is defined as ``the ability to sustain the coordinated deployment of assets in ways that help a firm achieve its goals''.

In this paper, `competence' is to be understood as defined by Gallon~\cite{article:galon:1995}:
\begin{quote}
a competence is an aggregation of capabilities, where synergy that is created has sustainable value and broad applicability.
\end{quote}

Several works on competence models have been published ~\cite{article:sanchez:1997,article:pepiot:2007}. Recently, the 4-C model, based on former models by ~\cite{article:Prahalad:1990, article:javidan:1998, www:hrxml, prove:molina:2000, article:mueller:2006, report:boucher:2005}, has been proposed by Ermilova and Afsarmanesh~\cite{article:ermilova:2010}. The 4-C model is adapted to characteristics and needs of VOBE, its members and VOs. The main components in the 4C-model are: Capabilities, Capacities, Costs, and Conspicuities. 

A first limitation of existing competence models is the lack of support for circumstantial and multi-version competences. In the context of SOVOBEs, a second limitation of competence models is the insufficient support of both VO-related and service-oriented concepts. The 4-C model---which has been developed to model competences in VOs---supports VO-related concepts, but lacks support for service-oriented concepts.

The description of the competences of an organization is usually complex because of the diversity and multi-aspect character of competences. Additionally, the continuous adaptation of SOVOBE members to market needs causes a significant effort related with the maintenance of this information. Therefore, the amount of information concerning the competences of VOBE members is significant, especially in medium and large SOVOBEs. As a consequence, computer support for management of competences is required in medium and large SOVOBEs, usually based on a \emph{competence model}.

The competence model is usually an important element of tools provided by VOBEs to support VO partner selection during the VO creation process ~\cite{book:lcm:2008}. An approach to VO partner selection based on information available in competence model is called \emph{competence-based configuration of VO} or \emph{competence-based VO creation}~\cite{article:ermilova:2010}.

\section{Introducing the proposed competence model}
\label{sec:introducingTheProposedCompetenceModel}

\subsection{Rationale}

In SOVOBEs, the partner selection process is directly connected with service selection, partner competences being considered as an extension of service description. There is currently a multiplicity of approaches to service description elaborated in isolation from existing competence models. Proposed service and competence description models do not specify the actual relation among competences and services. The definition of the relation among these concepts is crucial for partner and service selection in SOVOBEs based on both competence model and service description.

Furthermore, many elements that are traditionally part of competence description models, \eg organization costs and capacities, in particular availability of resources, depend on circumstances (\eg seasons, days, economic environment, client's country of origin). Additionally, as organizations evolve in time, so should the description of their competences. Therefore, competence description models should support circumstantial and multi-version competences. 

Existing competence models do not deal neither with circumstantial and multi-version competences, nor with service orientation. Even the 4-C model which is the closest to the needs of SOVOBEs, has to be refined and extended to support SOVOBEs.

As a consequence, the shift of VOBEs to the SOA paradigm observed in SOVOBEs leads to the development of novel competence models adapted to SOVOBEs, supporting the characteristics of both SOA ecosystems and VOBEs.

\subsection{Overview of approach}
The presented model may be considered as a refinement of the 4-C model taking into account the service orientation of SOVOBEs. 

A competence model should provide exhaustive information about an organization, its competences, and the services it provides. A competence model should be paired with  methods to verify the reliability and relevance of competence descriptions. Finally, it should be possible to search organizations whose profiles are stored in a competence description model. Therefore, the proposed \emph{competence model} consists of:
\begin{itemize}
 \item a competence description model,
 \item a competence verification method, and
 \item a competence search method.
\end{itemize}

The \emph{competence description model} is an information model for  comprehensive descriptions of various aspects of organization operations, \eg information about organization competences, utilized resources, production capacities, financial and legal issues, ownership, employees. An important part of an organization description is the business characteristic of the services provided by the organization, and the relations between services and competences needed to provide them. This relation links SOA to competence modelling, formerly developed separately. Information concerning competences and services is provided by the organization itself and is stored in a competence repository for later access by SOVOBE members, and potentially external organizations.

For competence descriptions of an organization to be reliable and relevant, it should be confirmed or \emph{verified} against other sources of information. In this context, various sources of information available in SOVOBE are proposed for this purpose. In particular, a social network may encompass information concerning social relations among organizations. These relations may, in turn, refer to the history of collaboration among organizations, and recommendations and opinions of SOVOBE members about other organizations. 

Both the competence description model and the method for the verification of competence descriptions are a basis for the \emph{competence search method}. The competence search method is based on the concept of competence requirement, being a set of competence properties and predicates expressing expected value of these properties. The competence search method encompasses the definition of competence requirements and evaluation of particular organizations against them. The competence search method in this form may be incorporated into more complex and comprehensive method of partner and service selection for VOs. 

A competence model consisting of these three elements may be provided by a SOVOBE to its members and its environment as a service. The proposed model has been implemented in the \ergo system presented in more details in Section~\ref{sec:ergo}.

\section{Competence description model}
\label{sec:competenceDescriptionModel}

The competence description model consists of three types of profiles:
\begin{itemize}
	\item competence profile,
	\item service business profile, and
	\item organization profile.
\end{itemize}

\subsection{Competence profile}

A competence profile is organized around five main concepts: \emph{competence}, \emph{capability}, \emph{capability variant}, \emph{capacity} and conspicuity. 

These concepts are directly linked to the concepts of \emph{service} and \emph{activity} that are a part of a \emph{service profile}. 

To our best knowledge, among all the proposed competence models, the 4-C model is the closest to the needs of SOVOBEs. However, the refinement of already proposed concepts and the introduction of new ones are still needed for the 4-C model to support SOVOBEs. Newly introduced concepts are presented in Table~\ref{tab:coreConcepts}. Concepts existing in the 4-C model by refined or redefined for SOVOBEs are presented in Table~\ref{tab:conceptsFrom4CModel}. Finally, concepts added to the 4-C model and associated with contextual competences are presented in Table~\ref{tab:newConcepts}.

\subsubsection{Core concepts}
\label{sec:coreConcepts}

The core concepts of the proposed model are \emph{activity}, \emph{service }(cf. Table~\ref{tab:coreConcepts}), \emph{competence}, and \emph{capability }(cf. Table~\ref{tab:conceptsFrom4CModel}).

While the \emph{service} is a type of a \emph{product} in the 4-C model (cf. Table~\ref{tab:conceptsFrom4CModel}), the proposed model is based on a different approach to \emph{service} concept, leading to significant changes in the relations among the concepts of \emph{service}, \emph{activity}, \emph{competence} and \emph{product}. An activity (a piece of work that delivers a certain product) can be performed by an organization presenting the associated capability (the ability to perform an activity). A competence aggregates one or more capabilities, and eventually other competences. A service is a mechanism to provide external organization with an access to competences.

\begin{figure*}[!t]
\centering
\includegraphics[width=6.0in]{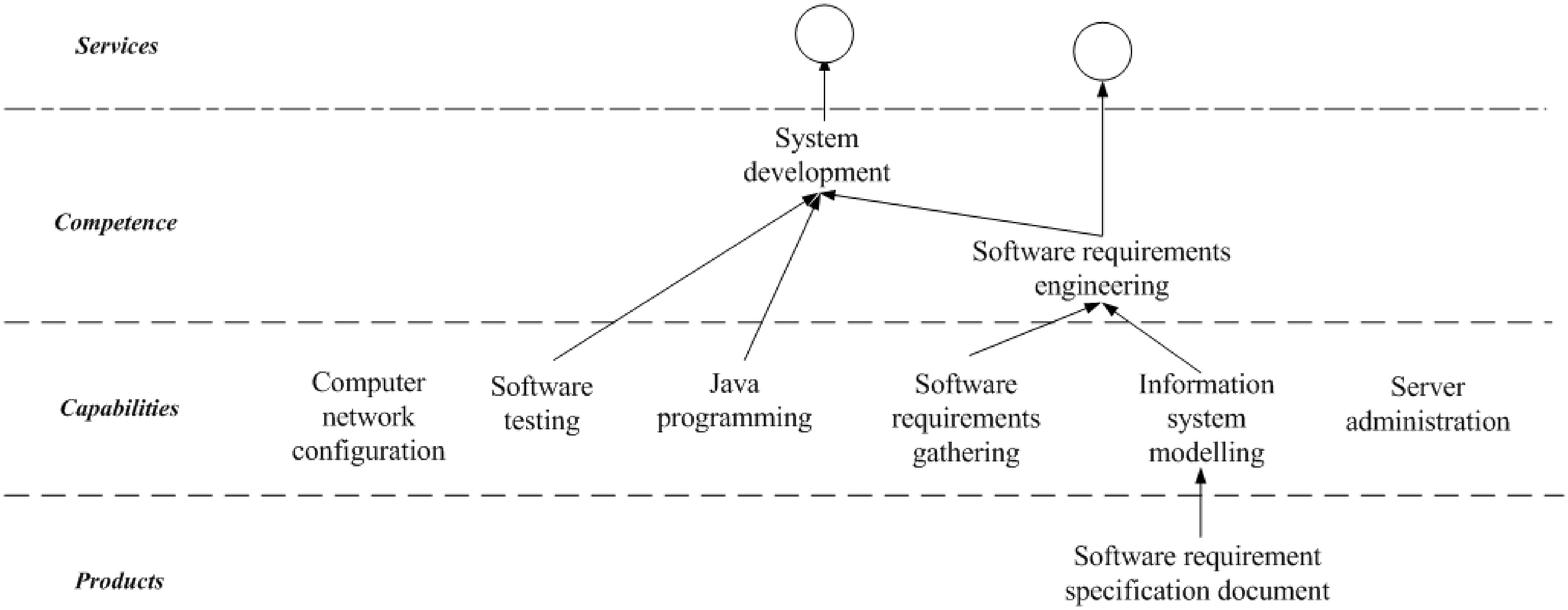}
\caption{An example of a competence profile}
\label{fig:CompeteceExample}
\end{figure*}

\sloppy
As an example, consider a software company (cf.~Figure~\ref{fig:CompeteceExample}). The company has a number of capabilities, for instance \texttt{server administration}, \texttt{computer network con\-fi\-gu\-ra\-tion}, \texttt{in\-for\-mation system modelling}, \texttt{software re\-quire\-ments gathering}, \texttt{Java programming}, \texttt{software testing}. Each capability is associated with an activity that results in a product: for instance the software company is capable of performing the \texttt{software re\-quire\-ments gathering} activity resulting in the product \texttt{soft\-ware requirement spec\-i\-fication document}. Ca\-pac\-i\-ties may further be aggregated into competences, \eg capabilities \texttt{information sys\-tem modelling}, \texttt{soft\-ware re\-quire\-ments gathering} are aggregated into the \texttt{soft\-ware requirements engineering} competence. A competence may also consist of competences: the competence \texttt{system development} is a compound competence composed of the \texttt{software requirements en\-gi\-neer\-ing} competence, and the \texttt{Java programming} and \texttt{software testing} capabilities. Finally, a competence may be externalized with an appropriate service: the \texttt{software requirements en\-gi\-neer\-ing} competence may be externalized as a service that may be then consumed by customers.
\fussy

Note that some capabilities described in competence description model may not be externalized by an organization as a part of a competence. The description of capabilities referring to activities that are not externalized is justified, \eg for private, internal enterprise architecture modelling, for collaboration opportunity spotting, or competence gap analysis~\cite{article:ermilova:2010}.

\begin{figure*}[!t]
\centering
\includegraphics[width=7.2in]{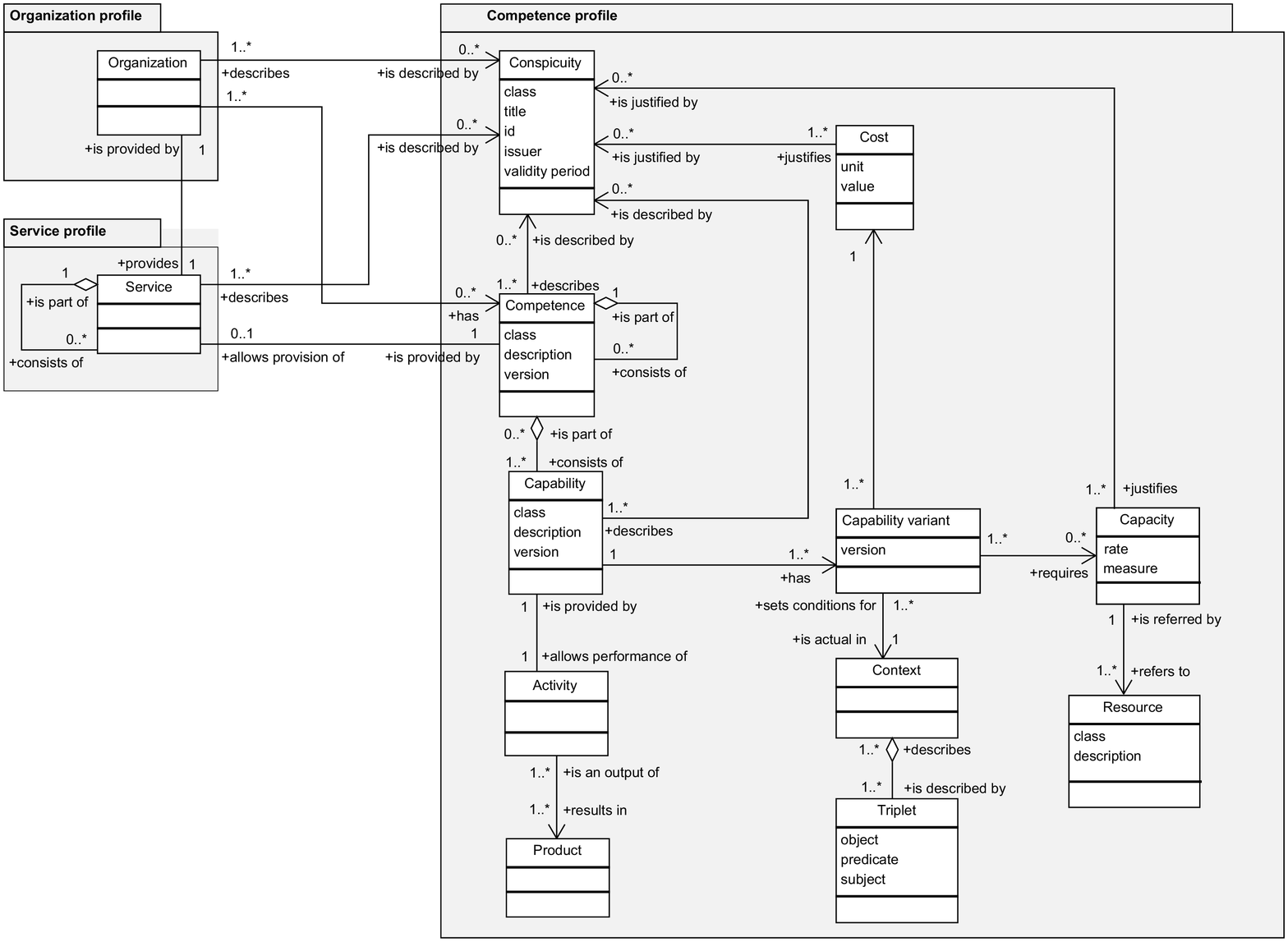}
\caption{Competence Description Model}
\label{fig:UML}
\end{figure*}

Organization's competences are in 1:1 relation with organization's services meaning that every competence is externalized by none or exactly one service (cf. Figure~\ref{fig:UML}). Over time, organization may provide more complex services, created not only as a composition of activities the organization is able to perform due to its capabilities, but also as an aggregation of other services. Such services are called \emph{compound services} (cf. Table~\ref{tab:coreConcepts}). Similarly, competences may be aggregated into \emph{compound competences} (cf. Table~\ref{tab:coreConcepts}) to provide more complex, possibly compound services. 

\begin{table}[!t]
\renewcommand{\arraystretch}{1.3}
\caption{Core concepts of the proposed competence model}
\label{tab:coreConcepts}
\centering
\begin{tabular}{p{1.5cm}p{6.5cm}}
\toprule
\bfseries Concept & \bfseries Proposed model\\
\midrule
Activity	& A piece of work that forms one logical, self-contained whole. The output of an activity is a product. An activity may be a manual activity or automated and requires human and/or machine resource(s) to support its execution~\cite{standard:WFMC:terminology}.
As stated in~\cite{standard:WFMC:terminology}, a task is a synonym of activity.\\
Service & A mechanism to enable access to one or more capabilities, where the access is provided using a prescribed interface and is exercised consistent with constraints and policies as specified by the service description~\cite{standards:OASIS:SOA:referenceArchitecture}. An access to a set of capabilities is possible with the concept of a competence. \\
Compound service & A service that is an aggregation of services. 
Aggregation of services creates new, more complex service.\\
Compound competence & A competence that is an aggregation of competences.
Complex competences may be defined as aggregation of other competences.\\
Product & The output of the activity.\\
\bottomrule
\end{tabular}
\end{table}

In Table~\ref{tab:conceptsFrom4CModel}, definitions proposed in 4-C model are presented in column ``Definition'', with comments on the proposed definitions given in the column ``Comment''. The refined or redefined definitions are presented in the column ``Proposed model''.

\begin{table*}[!t]
\renewcommand{\arraystretch}{1.3}
\caption{Refined or redefined concepts from the 4-C model}
\label{tab:conceptsFrom4CModel}
\centering
\begin{tabular}{p{1.5cm}p{4cm}p{5.5cm}p{5.5cm}}
\toprule
\multirow{2}{*}{\parbox[c]{1.5cm}{\centering \bfseries Concept}}	& \multicolumn{2}{c}{\bfseries 4-C model} & \multirow{2}{*}{\parbox[c]{5.5cm}{\centering \bfseries	Proposed model}} \\ \cmidrule{2-3}
&	\parbox[c]{4cm}{\centering \bfseries	Definition} & \parbox[c]{5.5cm}{\centering \bfseries Comment} & \\
\midrule
Competence & Competency is a compound object that cannot be represented by one textual value. & Too general definition & An aggregation of capabilities, where synergy that is created has sustainable value and broad capability~\cite{article:galon:1995}.\\
Capability & An ability to perform an activity or task. & Missing discussion on a difference among ``task'' and ``activity''. According to~\cite{standard:WFMC:terminology}, ``task'' is a synonym of ``activity''. & An ability to perform an activity.\\
Cost & Represent the cost of product/services provision in relation to one capability. & Missing clear definition &	The monetary value of all the expenditures linked to activity addressed by particular capability, including the value of all the resources required by an activity.\\
Resource & Resource class represents the elements applied to business processes in the organization. & Missing clear definition & Physical or virtual entity of limited availability required by organization to perform activities and achieve organizational goals.\\
Capacity & The current availability of resources needed to perform one specific capability. & In this model, a capability is defined as ``an \emph{ability} to perform an activity'', resulting in ``resources needed to \emph{perform} one specific \emph{ability} to perform activity'' - the expression ``performing an ability'' makes this definition unclear. & The total amount of product that can be contained or produced.\\
Conspicuity	& Represent means for the validity of information provided by the VOBE members about their capabilities, capacities and costs. & Missing clear definition.\newline
Refers only to: capability, capacity and cost. 	& A formal or informal document justifying, confirming and explaining information provided in a competence description.\newline
Refers to: service, organization, competence, capability, cost, capacity.\\
Product & Represents both goods and services that belong to the output of the processes/activities represented by the member organizations' capabilities. & Missing clear definition.
\newline
Assumed in the proposed model concentration around the \emph{service} concept requires the redefinition of the \emph{product}. & The output of the activity.\\
\bottomrule
\end{tabular}
\end{table*}

Concepts presented above and relations existing among them are presented in Figure~\ref{fig:UML} in a form of UML diagram.
The concepts presented in Figure~\ref{fig:UML} are grouped in organization, service, and competence profiles. Note that organization and service profiles may be extended as needed.

\subsubsection{Contextual capabilities}

In addition to terms defined in 4-C model, the proposed competence profile introduces the concepts of \emph{capability context}, \emph{capability variant} and \emph{versioning} (cf. Table~\ref{tab:newConcepts}).

Conditions under which an organization is able to perform some activity depend on circumstances. Those circumstances are referred to as \emph{context}. Depending on the context, cost and capacity may vary. This results in a number of \emph{capability variants} of particular capability. A capacity variant consists of a particular capability context, a particular capacity, and a particular cost. Moreover, the model includes versioning of competences, capabilities and capability variants. \emph{Versioning} allows for tracking of the evolution of an organization and its adaptation to market needs and collaboration opportunities.

Continuing the former example, the number of available \texttt{programmers} (considered here as resources) may be lower than usual in particular circumstances, such as \texttt{holi\-days}. This limitation results in different capability characteristics, for instance extended \texttt{system development time} or increased \texttt{general cost} of system development. These characteristics constitute a capability variant associated with a particular context (\texttt{holidays}).

\begin{table}[!t]
\renewcommand{\arraystretch}{1.3}
\caption{Concepts related with contextual capacities}
\label{tab:newConcepts}
\centering
\begin{tabular}{p{1.5cm}p{6.5cm}}
\toprule
\bfseries Concept & \bfseries Proposed model\\
\midrule
Capability context & A set of triplets $\langle$object, predicate, subject$\rangle$ describing circumstances in which a capability exists.\\
Capability variant & Cost and a set of capacities referring to particular capability and appearing in particular context.\\
Version & A number indicating a competence, capability or capability variant version\\
\bottomrule
\end{tabular}
\end{table}

\subsection{Service profile}

In SOA, various standards supporting Web service description, \eg WSDL~\cite{martin:2005}, OWL-S~\cite{standards:DAML:OWLS}, WSMO~\cite{book:fensel:wsmo}, have been developed to provide information necessary to find a service and to interact with it. These standards enact service discovery, service invocation, service composition, and interoperation~\cite{standards:DAML:OWLS}. At a business level, the scope of relevant information that is included in service description should focus on \emph{business characteristics of a service}, without technical aspects, \eg service reference to organization strategic goals, strategy of service, formal requirements. ~\cite{prove:boukadi:2009,standard:togaf9}. A set of information relevant for service profiles has been partially developed within the ECOLEAD project~\cite{prove:ermilova:2006}.

A service provider can provide complex services according to its competences. Thus, in the proposed approach, a service is connected with a 1:1 relation with a competence. A service is defined as a mechanism enabling an access to a set of capabilities defined as a competence. 

\subsection{Organization profile}

The description of organization profiles should include information that are not specific to services, \eg membership in associations, localization, financial capital, contact information, and steering managerial board. A scope of information relevant of organization profile has already been developed within the ECO\-LEAD project~\cite{prove:ermilova:2006}.

\section{Competence verification method}
\label{sec:competenceVerificationMethod}

The information provided by an organization should be confirmed or verified against other sources of information about this organization to be reliable. The competence description model allows organizations for an initial verification of data reliability based on \emph{conspicuities} (cf. Table~\ref{tab:conceptsFrom4CModel}). 

In addition to conspicuities usually provided by a particular SOVOBE member, the SOVOBE itself stores various sources of information that can enable the verification of information provided in the competence description model. These sources of information may for instance offer access to information about the SOVOBE members, their history of collaboration, efficiency of collaboration, former and existing problems (cf.~Table~\ref{tab:dataSourcesForVerificationOfCompetences}).

\begin{table}[!t]
\renewcommand{\arraystretch}{1.3}
\caption{Examples of data sources for the verification of competence descriptions }
\label{tab:dataSourcesForVerificationOfCompetences}
\centering
\begin{tabular}{p{1.7cm}p{6.3cm}}
\toprule
\bfseries Data source & \bfseries Description\\
\midrule
Continuous monitoring of collaboration & Monitoring of current service consumption and provision, progress in running collaboration processes, conformance to requirements\\
History of collaboration & Information restored from the historical data about partners' performance and collaboration within SOVOBE\\
Opinions of SOVOBE members & Information provided by SOVOBE members concerning to other members' competences and services\\
Repository of social relations & Information about relations among organizations (\ie recognition, trust) available in SOVOBE's social network~\cite{prove:swierzowicz:2009,prove:picard:2009}\\
\bottomrule
\end{tabular}
\end{table}

Information concerning the history of collaboration, opinions of SOVOBE members, and their social relations can be stored in a social network, either by VOBE members themselves, or by third parties and information systems monitoring the operations of organizations (including VOs). A social network consists of a set of individuals and organizations linked by ties. Social networks are often represented by graphs, with nodes representing individuals and organizations, and arcs representing ties~\cite{article:swierzowicz:2009}.

A social network may be considered as a complementary source of information that significantly contributes to the meaning and usefulness of information stored in the competence description model. During competence search, an analysis of the competence model information should be complemented by an analysis of the social network. For example, one organization may claim to have many successful projects in collaboration with other VOBE members, although the same organization is connected with a very limited number of VOBE members in the social network. Additionally, the nature of the relations in the social network may indicate that the organization has a low trust and recognition level from other VOBE members, leading to the conclusion that, although the organization claim to be a successful organization in its competence profile, it should probably be avoided as a collaborator in future VOs.

The importance of social aspects in SOA has been noted recently ~\cite{article:jarimo:2009,prove:swierzowicz:2009,picard:prove:2009}. Models of social relations, models of requirements incorporating these relations and techniques for their analysis are not yet mature. As an example, Ding and al.~\cite{article:ding:2003} have proposed a simulation-optimization approach using genetic search for supplier selection, integrating performance estimation, social aspects and genetic algorithm. However, the social relation model encompasses only a simple social model for supply chains limited to only one relation type, \ie customer-supplier.
\section{Competence search method}
\label{sec:competenceSearchMethod}

Competence search consists in determining in a searchable domain, the set of organizations satisfying a given set of expectations concerning their competences, referred to as competence requirements. In the context of competence search, the searchable domain is the set of organizations whose competences have been modelled and stored in the competence repository. Two elements of a search method---and, in particular, competence search method---may be distinguished.
First, an information model captures and structures information about elements of the searchable domain. Second, a search technique seeks appropriate information in the searchable domain according to a given search query. A review of various search techniques focusing on the search context, scope of the search query and search strategy may be found in~\cite{article:crispim:2007}. 

In our approach, the information model consists of a \emph{competence description model} and \emph{competence requirements model}. The Multi-Aspect Partner and Service Selection (MAPSS) method~\cite{picard:prove:2010} may serve as an example of a search technique. 

\subsection{Competence requirement model}

The competence model description presented above relies on a set of concepts to describe organizations operating within a SOVOBE. These concepts may be represented as sets of properties. In this paper, a set of properties is referred to as `object'.
This simple property representation is proposed as a basis for the definition of competence requirements:
\begin{quote}
a competence requirement is a property of an organization and its associated expected values, which expressed as a predicate. 
\end{quote}
A competence requirement is satisfied by an organization is the associate predicate is true for the value of the given property of the organization. The envisioned requirements may include a list of required competences, a list of required capabilities with a clear statement of required capacity and optimal cost in particular circumstances defined by a context, or a list of required conspicuities (\eg certificates, diplomas). A set of related competence requirements is referred to as \emph{organization class}. 
The competence requirement model relies on external mechanisms for the evaluation of competence requirements as regards a given organization.

As an example, consider an \emph{organization class} \texttt{Polish Software Company}, defined by the following set of competence requirements:
\begin{itemize}
 \item $\langle \texttt{organization:profile:localization},$ $= \texttt{Poland}\rangle$,
 \item $\langle \texttt{competence:name},$ $\supset \{\texttt{Java programming}\}\rangle$ ,
 \item $\langle \texttt{capability:name},$ $\supset \{\texttt{Server administration}\}\rangle$.
\end{itemize}
The first requirement concerns properties named \texttt{organi\-za\-tion:profile:localization}. A property satisfying the associated predicate $= \texttt{Poland}$ has to have its value equal to $\texttt{Poland}$.

Consider an \emph{organization} \SoftwareDev defined by the following set of properties:
\begin{itemize}
\item $\langle \texttt{organization:profile:name},$ \\
			\hspace*{.5cm}$\SoftwareDev\rangle$,
 \item $\langle \texttt{organization:profile:localization},$ \\
 			\hspace*{.5cm}$\texttt{Poland}\rangle$,
 
 \item $\langle \texttt{organization:profile:creationDate},$ \\
 			\hspace*{.5cm}$\texttt{Nov, 1st, 2009}\rangle$,
 \item $\langle \texttt{organization:profile:numberOfEmployees},$ \\
 			\hspace*{.5cm}$\texttt{34}\rangle$,
 \item $\langle \texttt{competence:name},$ \\
 			\hspace*{.5cm}$\{\texttt{Java programming},$  \\		
 			\hspace*{.5cm} $\texttt{Ruby programming},$ \\
 			\hspace*{.5cm} $\texttt{Python programming},$ \\
 			\hspace*{.5cm} $\texttt{Software requirements engineering}\}\rangle$ and,
 \item $\langle \texttt{capability:name},$ \\
 			\hspace*{.5cm}$\{\texttt{Server administration},$ \\
 			\hspace*{.5cm} $\texttt{Computer network configuration}\}\rangle$.
\end{itemize}

\begin{figure*}[!t]
\centering
\includegraphics[width=3.6in]{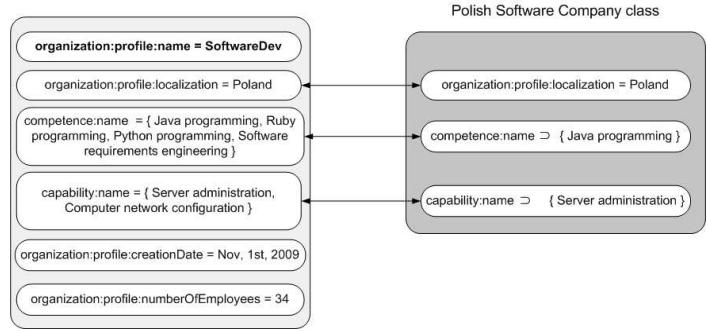}
\caption{An example of organizations which are instances of a class. All requirements are satisfied.}
\label{fig:RequirementSatisfaction}
\end{figure*}

All the competence requirements of the class are satisfied by \SoftwareDev (cf. Figure~\ref{fig:RequirementSatisfaction}). The organization \SoftwareDev is therefore an \emph{instance} of the class \texttt{Polish Software Company}.

In the presented example,  the default logical operator combining the requirements is the AND-operator. It is not always the case, as other logical operators may be used. Following on the former example, one may want to define an organizational class expressing the need for an organization operating in the market for at least two years and having a competence \emph{System development}, or, alternatively, an organization that has conducted at least ten software projects in the past three years.
 

Note that some properties of \SoftwareDev, such as  \texttt{organi\-za\-tion:profile:name}, \texttt{organization:profile:\-creationDate}, and \texttt{organization:profile:numberOf\-Em\-ployees}, are meaningless as regards the class \texttt{Polish Soft\-ware Company}. 

\begin{figure*}[!t]
\centering
\includegraphics[width=3.6in]{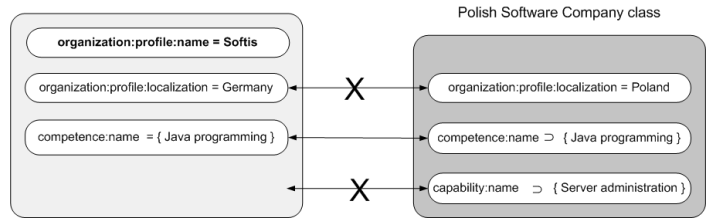}
\caption{An example of organizations which are \emph{not} instances of a class. Requirements concerning \emph{localisation} and \emph{capability name} are not satisfied.}
\label{fig:RequirementUnsatisfaction}
\end{figure*}

In Figure~\ref{fig:RequirementUnsatisfaction}, the organization \texttt{Softis} is presented. This organization is not an instance of the \texttt{Polish Software Company} class: the requirement concerning \texttt{organiza\-tion:\-pro\-file:localization} property is not satisfied by the value of the property \texttt{organization:profile:localiza\-tion}. Finally, no property \texttt{capability:name} is even defined for \texttt{Softis}.

\subsection{Multi-Aspect Partner and Service Selection in SO\-VO\-BEs}

The competence search problem may be considered as part of a larger \emph{partner and service selection problem}. Partner and service selection is a complex multi-aspect and multi-criteria process in which usually multiple parties are involved. This process can hardly be fully automated because of its complexity and the fact that it relies on information that are often tacit. In general, partner and service selection method encompasses the following aspects~\cite{picard:prove:2010}:
\begin{itemize}
 \item determination of roles in collaborative processes and corresponding requirements for organizations and services, as well as the planned structure of the VO to be created,
 \item identification of organizations and services able to play a particular role in VO collaborative processes,
 \item negotiation and settlement of collaboration rules and conditions,
 \item analysis of possible VO variants in terms of conformance to requirements and efficiency of collaboration.
\end{itemize}

In this context the Multi-Aspect Partner and Service Selection (MAPSS) method~\cite{picard:prove:2010} has been proposed as a method for selection of partners and services for VO collaboration processes. The MAPSS method supports social aspects, competences of VO members and performance characteristic. In the MAPSS method approach, each activity to be executed as part of a VO collaboration process is performed by a \emph{service consumer} consuming a \emph{service} provided by a \emph{service provider}. A \emph{process element} may refer to a service consumer, a service or a service provider. Service consumers and service providers are called \emph{partners}. In this context, a \emph{role} is a set of requirements that a potential partner has to satisfy to be assigned to a particular activity. Roles are in M:N relation with partners and services. 
The concept of \emph{role} in the MAPSS method is similar to the concept of \emph{organization class} being a part of competence description requirement model. The \emph{organization class} can be considered as an approach to implement the concept of roles. Competence requirements, in turn, can be considered as a part of VO specification.

Among requirements defining a role, social requirements constrain relations between roles. Social requirements may be used to define some properties of a social network and their associated expected values. A social (sub)network of organizations and services may then be evaluated against social requirements~\cite{article:swierzowicz:2009}. Examples of relations are past collaboration, recognition, and former financial exchange.  In context of this paper, social requirements are also important for the verification of information stored in the competence model. 

\begin{figure*}[!t]
\centering
\includegraphics[width=4.0in]{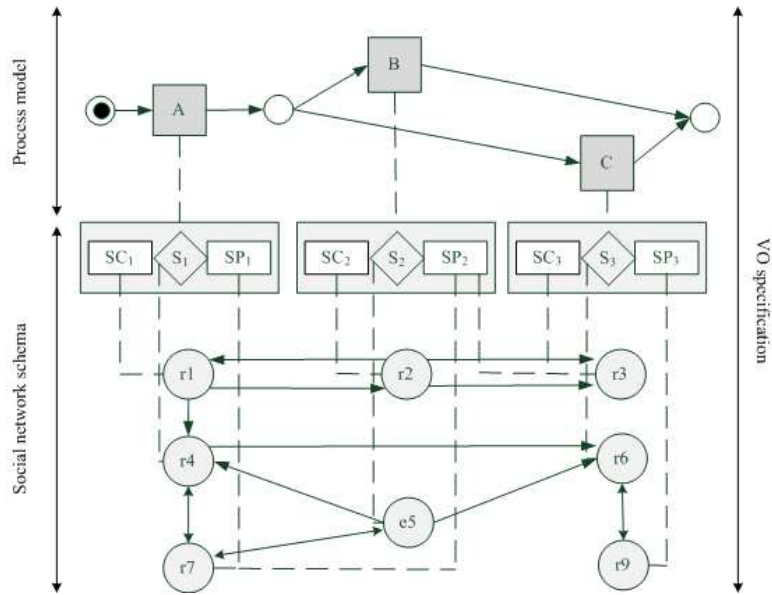}
\caption{VO specification for the MAPSS method.}
\label{fig:MAPSS}
\end{figure*}

The set of roles assigned to process elements and the relations among these roles are referred to as a \emph{social network schema}. A \emph{VO specification} consist of a process model, and associated requirements, paired with a social network schema (cf. Figure~\ref{fig:MAPSS}). 

For the proper selection of partners and services, the MAPSS method consists in five phases and follows the general selection method guidelines presented in~\cite{article:camarinhamatos:oliveira:2007}:
\begin{enumerate}
 \item definition of VO specification - definition of requirements and associated preferences (\eg importance, acceptable level of satisfaction);
 \item selection of partners and services for roles - the output of this phase is a set of partners or services for each role defined in the social network schema;
 \item VO variant generation - a variant is a group of organizations that may potentially be operating as a VO; the output of this phase is a sorted set of variants, according to the sorting preferences defined by a VO planner;
 \item performance evaluation - assignment of selected elements to process activities and validation of performance requirements;
 \item VO inception - registration of the newly created VO in the SOVOBE, especially in its competence and service repositories.
\end{enumerate}
In every phase, human action may lead to requirements redefinition, to modifications of the preferences, and to the repetition of phases of the MAPSS method.

One of the crucial modules of the method is a \emph{competence and service description module}. The competence and service description module is supposed to provide the following features for MAPSS:
\begin{itemize}
 \item structured description of organization competences,
 \item search of organizations based on competence-based competences,
 \item evaluation of the conformance of an organization to a set of defined competence-based requirements,
\end{itemize}

\section{Implementation concerns}
\label{sec:implementationConcerns}

The concepts described in this paper can used in various application areas and have been fully implemented as a part of the \ergo system.

\subsection{Potential applications}

The concepts described in this paper may applied to various business domains.

In the \emph{construction sector}, organizations are usually collaborating in a VO manner. In this sector, the general contractor, and sometimes the real-estate developer, plays the role of the VO planner, as well as the role of the SOVOBE itself: their role is to identify appropriate partners and services needed to construct a given building. The partner and service selection process is therefore of high importance for the general contractor.
 
Additionally, the concept of capability variants is particularly relevant in this sector. The availability of partner's free resources, associated with capability costs, capability properties such as time associated with capability, depend strongly on the context, \eg seasons of the year, weather conditions, day of week, hours, and holiday period. As an example, the performance of teams working on roofs is usually lower in winter than in spring because of weather conditions. Thus the inclusion of capability variants in the description of organization's competences is crucial for the proper selection of organization for a given business process in the construction sector.

A second sector in which the proposed solution may be applied is the \emph{healthcare sector}. In this sector, each hospitalized patient may be considered as a ``collaboration opportunity'' for the medical staff. Physicians, surgeons, and nurses share their competences to heal a patient. Hospital teams are usually formed dynamically to answer the needs of a given patient.

In this case again, the concept of capability variants is important to capture the availability of members of the hospital teams. Physicians, surgeons, and nurses often work at various places. Additionally, the work hour system in the healthcare sector is organized around the concept of duty hour, which leads to very flexible and changing schedules for the healthcare workers. Therefore, capability variants may support the partner search selection, under the condition that the schedules are translated into appropriate capability context.

A last example of a sector in which the proposed solution may be applied is the sector of \emph{humanitarian aid}. In this sector, the ``collaboration opportunity'' comes from natural disasters (\eg the 2011 Ha\"{i}ti earth quake or the 2004 Indian Ocean tsunami), technological disasters (\eg Tchernobyl or Fukushima Daiichi nuclear disasters), and long-term man-made disasters related to civil strife, civil war, and international war (\eg the 2011 Lybian conflict).

As a response to a given disaster, various organizations, usually non-governmental organizations (NGOs), collaborate to provide assistance to the population of the disaster region. Various competences are usually involved to respond to the emergency, \eg logistics to deliver food and drugs, or healthcare. An efficient response to a disaster requires the identification of the appropriate collaborating organizations. The proposed model may provide NGOs with support for disaster management and help them to organize aid in a more efficient and effective manner.

\subsection{The \ergo system}
\label{sec:ergo}

The competence description model, the competence verification method, and the competence search method (as part of the MAPSS method) proposed in this paper have been implemented in the \ergo system~\cite{www:ergo}.

\subsubsection{System outline}

The \ergo system is an implementation of a SOVOBE infrastructure. It allows a SOVOBE to provide its services for members or organizations in its surrounding environment. Although the number of internal services that a VOBE may offer to its members is theoretically unlimited, a set of internal services common to all the VOBEs, \ie \emph{core internal services}, may be identified. The \ergo platform is organized around five core internal services supporting the management of either VOBE members, or VOs. First, two core internal services focus on the management of SOVOBE members: \emph{the competence management service} provides means for structured description of SOVOBE members, while \emph{the social network service} addresses relations among SOVOBE members. Second, three core internal services focus on the management of VOs: \emph{the VO creation service}, \emph{the VO collaboration service}, and \emph{the VO monitoring service}. On a basis of core internal services, a SOVOBE tailored to the needs of a particular group of cooperating organizations or a particular sector may be built. A first pilot is currently under testing in the construction sector, with a real-estate development company being a SOVOBE, providing the infrastructure and services for its subcontractors on the \ergo platform.

A detailed description of the system components can be found on \ergo website ~\cite{www:ergo}, including the \emph{software requirements specification}, \emph{user's guide} and \emph{developer's guide} documents.

\subsubsection{Competence description model and competence search in the \ergo system}

Four services and modules of the \ergo system implements the functions related with the proposed competence description model and competence search:
\begin{itemize}
 \item VO collaboration service - implementation of the functions related with the definition of VO specifications, \ie the definition of a set of activities to be performed and of roles by grouping social requirements into classes;  
 \item competence management service - implementation of the competence description model, competence requirement definition and evaluation method and a mechanism for the partial verification of self-declared in partner selection process with the use of performance indicators~\cite{prove:paszkiewicz:2009};
 \item VO creation service - implementation of the MAPSS method including a mechanism for the verification of self-declared organization competences with the use of a social network;
 \item social network service - the repository of information used for verification of information stored in competence management service.
\end{itemize}

\begin{figure*}[!t]
\centering
\includegraphics[width=6.0in]{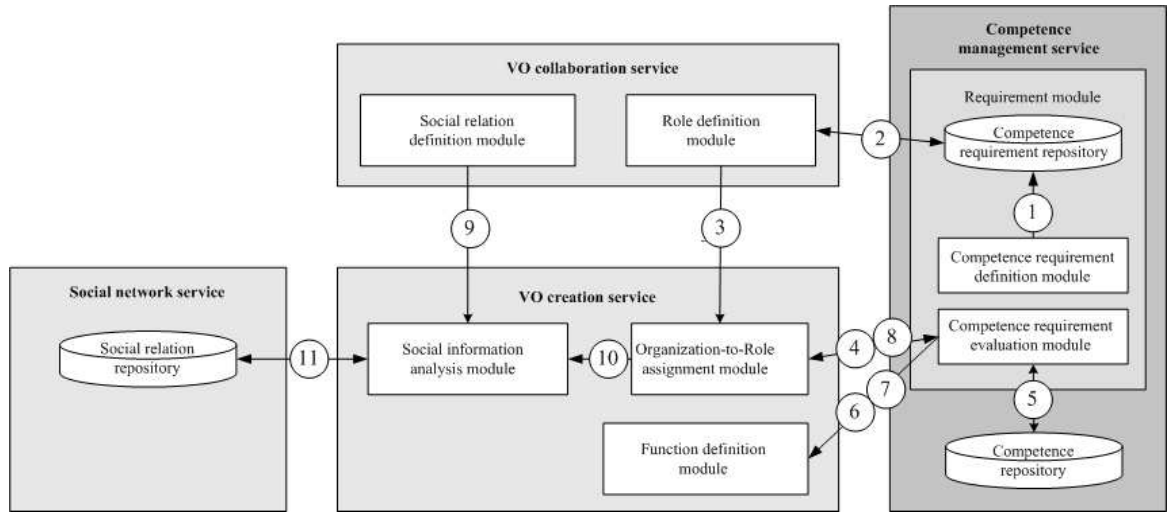}
\caption{Sequence of actions during VO specification creation and partner search.}
\label{fig:Communication}
\end{figure*}

In Figure~\ref{fig:Communication} the sequence of actions related with the definition of VO specifications and partner search in the \ergo system is presented. Only services and modules participating in this scenario are presented.

First, a set of competence requirements is defined. The competence requirement module is part of the competence management service, as the possibility of defining particular requirements depends on the information stored in the competence description model. Specified requirements are stored in the \emph{competence requirement repository} for future competence (partner) search processes (step~1 in Figure~\ref{fig:Communication}). During the creation of a VO specification, competence requirements are grouped into organization classes and linked with roles specified in social network schema (step~2). The VO specification serves as an input for the selection of partners and the search of competences performed in the \emph{Organization-to-Role assignment} module of the \emph{VO creation service} (step~3). This module takes advantage of requirements to specify a non-empty set of organizations that can perform a particular role. To this purpose, the competence requirements are sent to the \emph{competence requirement evaluation module} (step~4). This module evaluates organizations described in the \emph{competence repository} as regards competence requirement satisfaction (step~5). In this step, evaluation of the satisfaction of each requirement by particular organizations may be performed with a weighted function with different weights associated with requirements (step~6). The value of the function is returned to \emph{competence requirement evaluation module} (step~7) to order the set of organizations returned to VO creation service (8). So far, the selection is based exclusively on information stored in the competence description model. In steps 9, 10, and 11, organization competences are verified against the social network. The process of partner selection does not end at this point, but moves to the third phase of the MAPSS method. 

Communication among modules implementing functionality of services is based on events. This allows seamless addition and integration of new modules that can provide substitutive or complementary functionality. For instance there can be more than one module providing the functionality of definition and validation of competence requirements. It is up to the user which modules will be used.  Such seamless integration is a prerequisites of adaption of the \ergo system to the needs of particular SOVOBE.

\begin{figure*}[!t]
\centering
\includegraphics[width=6.0in]{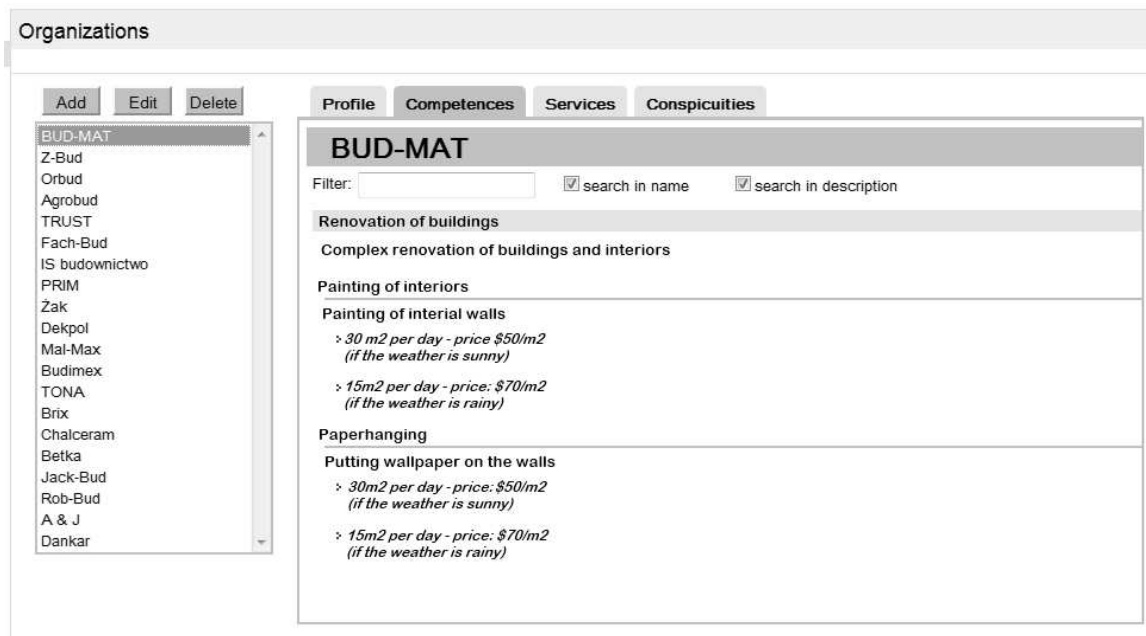}
\caption{Example screenshot from the ErGo system competence module.}
\label{fig:Screenshot}
\end{figure*}

In Figure~\ref{fig:Screenshot}, a screenshot of the \ergo system addressed presents the graphical user interface of the competence module allowing SOVOBE members to browse organization profiles, competences, services, and conspicuities.

\subsubsection{Programming platform}

The \ergo system is implemented in the Java SE 6 platform. The functionality provided by the system modules is exposed as OSGi services~\cite{www:osgi} within the Equinox container~\cite{www:equinox} and can be externalized as SOAP web services. The core frameworks used in the \ergo system are:
\begin{itemize}
 \item OSGi,
 \item Google Web Toolkit,
 \item Hibernate.
\end{itemize}
Components integrated within the \ergo system in the OSGi Container Equinox 3.5.2 are:
\begin{itemize}
 \item Hibernate 3.3.1.GA - communication with Oracle 11g;
 \item OSGi 4.2 Blueprint Container - OSGi services management;
 \item Jetty 6.1.19 - web application server;
 \item Google Web Toolkit 2.0.3 with various connected libraries - AJAX Web user interfaces;
 \item Apache CXF Distributed OSGi 1.2 - externalization of web services.
\end{itemize}

\section{Conclusions}
\label{sec:conclusions}

The main objective of this paper is to present the design and use of the competence model for SOVOBE members. The proposed competence model is composed of a competence description model, a competence verification method, and a competence search method (including a competence requirement definition technique). The proposed competence model largely extends the 4-C model. The proposed concept of competence profile clarifies the relation between activities, services, capabilities and competences, as well as introduces contextual capabilities.

The proposed description of competences takes into account the characteristics of SOVOBEs, redefining the concepts of \emph{competence}, \emph{capability}, \emph{resource}, \emph{capacity}, \emph{cost}, \emph{resource}, \emph{conspicuity}, and \emph{product}, and introducing the new concepts of \emph{capability context}, \emph{capability variant}, \emph{compound competence}, and \emph{version}. Most notable aspects of the proposed model include: 
\begin{enumerate}
\item modelling a \emph{context} for capabilities resulting in many possible \emph{capability variants}, 
\item versioning of competences, capabilities and capability variants, 
\item capturing relations among \emph{competence}, \emph{capability}, \emph{activity} and \emph{service}, with special emphasis on a clear distinction between the concept of competence and capability in context of service provision, 
\item modelling the multiplicity of relations among all the concepts of the proposed competence description model, 
\item the role of the social network in verification of information stored in competence description model, 
\item competence requirement model, and
\item putting competence model in context of partner selection in SOVOBEs.
\end{enumerate}

Among future works, a method for competence aggregation, suited to the proposed model, is to be proposed. Such a method should take into account the contextual and versioning aspect of capabilities, for a group of organizations. A major application could be the evaluation of the competences of a VO.

Another area of improvement concerns the evolution of competences in time. The proposed competence model should encompass strategies for the evaluation of competences in a more dynamic manner, assuming that an organization may actually have a competence at a given point in time, but at another point of time, the competence may not be available anymore.

Finally, the on-going deployment of the \ergo system in the construction sector will allow for an evaluation of the pertinence of the proposed competence description model.



\bibliographystyle{model1-num-names}
\bibliography{biblio}

\end{document}